\documentclass[twocolumn,showpacs,preprintnumbers,amsmath,amssymb]{
revtex4}
\usepackage{amsmath,amssymb,amsfonts,graphicx}
\usepackage{graphicx,epsfig}

\newcommand{\avg}[1]{\langle #1\rangle}
\newcommand{\req}[1]{Eq.~(\ref{#1})}

\newcommand{\fig}[1]{Fig.~\ref{#1}}
\newcommand{\tab}[1]{Table \ref{#1}}

\usepackage{color}

\def\cO{{\cal O}}
\def\cP{{\cal P}}

\def\cP{{\cal P}}

\def\cX{{\cal X}}

\def\anc{{\alpha}}
\def\anctwo{{\alpha'}}
\def\peer{{i'}}
\def\simi{{\em SIM} }
\def\rand{{\em RAN} }
\def\lon{{\em LON} }

\begin{document}

\preprint{}

\title[Title]
{Tracing the Evolution of Physics on the Backbone of Citation
Networks}

\author{S. Gualdi$^1$, C. H. Yeung$^{1,2}$, Y.-C. Zhang$^{1,3}$}
\affiliation{$^1$Department of Physics, University of Fribourg,
CH-1700
Fribourg, Switzerland\\
$^2$The Nonlinearity and Complexity Research Group, Aston University, 
Birmingham B4 7ET, United Kingdom\\
$^3$ Web Sciences Center, School of Computer Science and Engineering,
University of Electronic Science and Technology of
China, Chengdu 610054, P. R. China}

\date{\today}

\begin{abstract}
Many innovations are inspired by past ideas in a non-trivial way.
Tracing these origins and identifying scientific branches is crucial 
for research inspirations. In this paper, we use citation relations 
to identify the \emph{descendant chart}, i.e. the family tree of 
research papers. Unlike other spanning trees which focus on cost or
distance minimization, we make use of the nature of citations and
identify the most important parent for each publication, leading
to a tree-like backbone of the citation network.
Measures are introduced to validate the backbone as the descendant chart.
We show that citation backbones can well characterize the
hierarchical and fractal structure of scientific development,
and lead to accurate classification of fields and sub-fields.
\end{abstract}


\maketitle


\section{Introduction}

Many innovations are inspired by past ideas in a non-trivial way. 
Examples in statistical physics include the connection between spin
glasses and combinatorial problems \cite{mezard87, nishimori01}, the
application of critical phenomenon in earthquake modeling \cite{bak02,
corral04}, and the analyses of disease spreading by percolation theory
\cite{newman02, dorogovtsev08}. 
To draw these connections is easy, but to map individual fields onto a
\emph{descendant chart}, i.e. a family tree of research branches is
more complicated. An even more difficult task is to uncover the
macroscopic tree based on the microscopic relations between
publications. Despite the difficulties, the descendant charts are
crucial for revealing the non-trivial connections between branches
which stimulates inspirations. Accurate descendant charts also give a
natural classification of papers. 

A solid basis to study descendant charts is represented by the
citation network which can be seen as the original map of scientific
development.
In recent years, the citation and authorship networks have been
used to evaluate the impact of academic papers and scientists
\cite{chen07, radicchi09}. Though useful informations are retrieved,
most studies focus on contemporary impact and ignore the intrinsic
hierarchical structure of citations encoding the generation of
scientific advances. Unlike the horizontal exploration in
conventional paper classifications \cite{griffiths04}, we explore the
vertical, i.e. temporal, dimension in citation networks to 
identify the descendant charts of publications.

At this end we identify a \emph{backbone} of the citation network
by removing all but the most relevant citation for each paper.
The backbone hence results in a tree-like structure and is found
solely based on citation relations with no additional information.
Similar concepts of spanning trees are extensively studied in
transportation networks and
oscillator networks, as minimum spanning trees in terms of traveling
cost \cite{aldous06, jackson10}, and trees that maximize betweenness
\cite{goh06} or synchronizability \cite{nishikawa06}. Though the
citation backbone can be constructed by these definitions, we see no
direct correspondence between them and scientific descendant trees.
Instead, one should make use of the nature of citation relations and
identify the important parent and thus the offspring for each paper,
which constitutes a backbone specific for citation networks.

In this paper, we identify the descendant chart for publications in
journals of
American Physical Society (APS), based on their citation
network from year 1893 to 2009. Our objectives are three-fold.
Firstly, we introduce a potential approach to identify the most
relevant parent (among the set of original references) for
each publication which leads to a backbone of the citation
network. Secondly, we introduce measures to validate the
citation backbones as representative descendant charts and compare our
approach with two other simple procedures (i.e. selection of random
parent or the longest path to the root).
Finally, we show that citation backbones possess
features of hierarchy and self-similarity, and lead to a valid
classification
of papers in linear-time, compared to conventional polynomial-time
algorithms \cite{hastie01, fortunato10}. The present work pinpoints
the importance of scientific descendant charts, as well as their
intrinsic difference from other spanning trees.

\section{Methods}\label{methods}

\begin{figure}
\center
\includegraphics[scale=0.33]{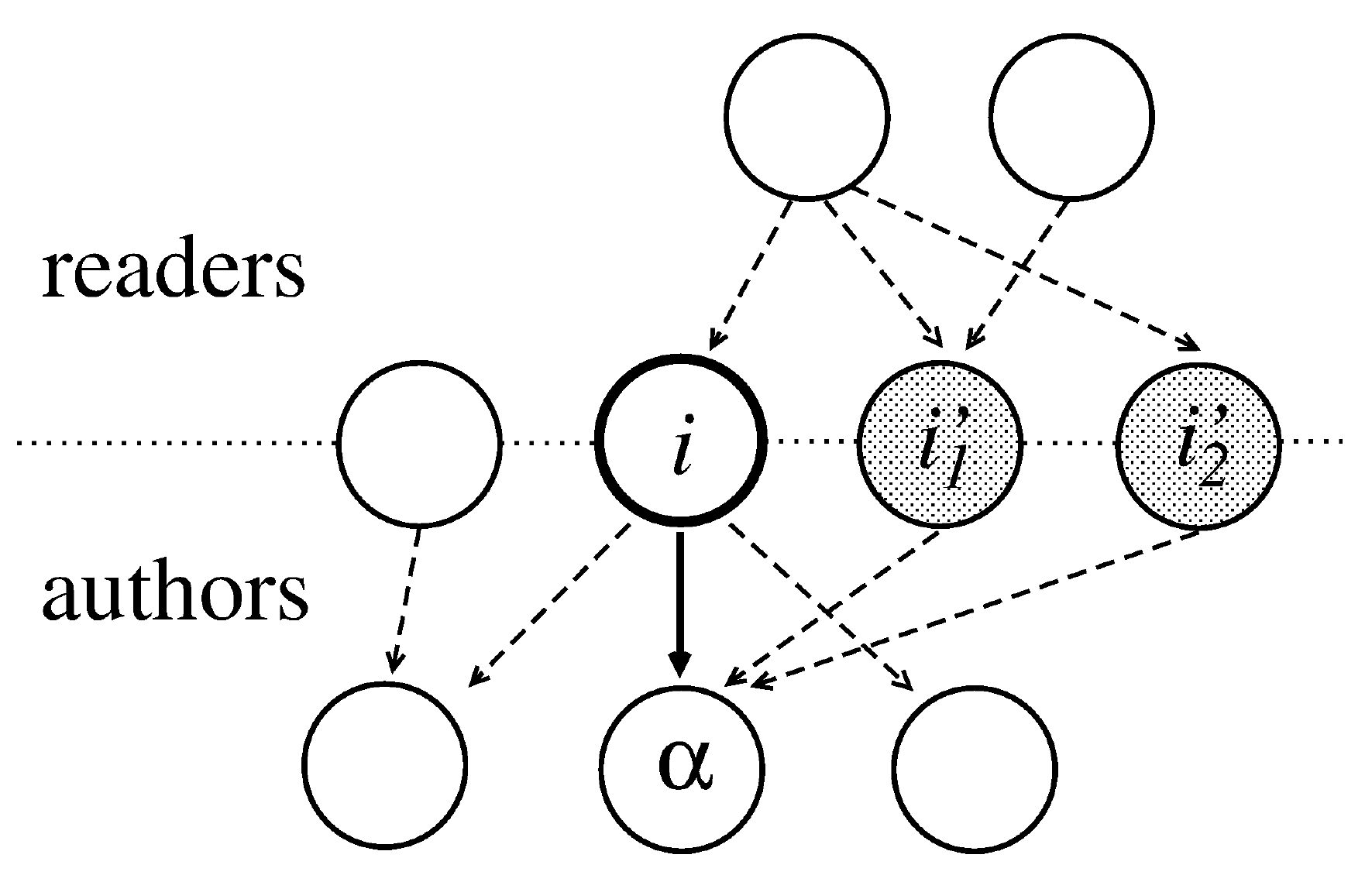}
\caption{A schematic diagram which shows two peers $i'=i'_1,i'_2$
(shaded) of $i$ rooted from parent $\anc$. To compute
each $s_{ii'}^{\rm aut}$ we consider a random walk from $i'$
through papers cited by both $i'$ and $i$. Specularly, 
to compute each $s_{ii'}^{\rm read}$ we consider a random walk from
$i'$ through
papers citing both $i'$ and $i$.}
\label{fig_tree}
\end{figure}

To start our analyses, we first denote the references of a paper as
its \emph{parents}, and the articles citing the paper as its
\emph{offspring}.
The set of parents and the offspring of a paper $i$ are denoted 
by $\cP_i$ and $\cO_i$ with  respectively $p_i$ and $o_i$ elements.
Intuitively, the offspring of an important paper should share
similar focus introduced by its influential parent.
Less relevant parents will by contrast lead to a more heterogeneous
descendance.
We thus quantify the \emph{impact} of parent $\anc$ on $i$ by
$I_{\anc\rightarrow i}=\sum_{i'\in\cO_\anc\backslash \{i\}}s_{ii'}$ 
where $s_{ii'}$ is the \emph{similarity} between $i$ and $i'$.
We refer to papers in the set of
$\cO_\anc\backslash \{i\}$ as the \emph{peers} of $i$ rooted in
$\anc$ (see \fig{fig_tree} for an illustration).
The higher the overall similarity between $i$ and the papers in
$\cO_\anc\backslash \{i\}$, the higher the impact of $\alpha$ on $i$.

A simple way to measure the similarity between $i$ and peer $i'$ is
to count the number of their common references, i.e.
$s_{ii'}=|\cP_i\cap\cP_{\peer}|$.
However, this similarity measure favors peers with many references,
resulting in an impact biased towards parents with a large
offspring. 
This suggests to define a  similarity measure based
on a random walk from the peers to $i$. We thus consider a
2-step random walk from each peer $i'$ to $i$ which passes
through their common references, and define a
contribution to $s_{ii'}$ as
\begin{eqnarray}
\label{eq_au}
s_{ii'}^{\rm aut}=\frac{1}{p_{\peer}}\sum_{j\in\cP_i\cap\cP_{
\peer} } \frac { 1
}{o_j}.
\end{eqnarray}
The superscript represents the \emph{authors'} interpretations,
as this similarity is measured through the references chosen by the
authors of $i$.
A second contribution is instead
given by a random walk through articles citing $i$ and represents the
\emph{readers'} interpretation of $i$. In analogy with $s_{ii'}^{\rm
aut}$ we define
\begin{eqnarray}
\label{eq_re}
s_{ii'}^{\rm read}=\frac{1}{o_{\peer}}\sum_{j\in\cO_i\cap\cO_{
\peer}}\frac{1}{p_j}.
\end{eqnarray}
As defined in both Eqs. (\ref{eq_au}) and (\ref{eq_re}), the higher
the random walk probability from $i'$ to $i$, the higher the
similarity between $i'$ and $i$. 

Finally, by combining linearly $s_{ii'}^{\rm read}$ and
$s_{ii'}^{\rm aut}$ and summing over all the peers rooted in
$\anc$, we obtain the impact of $\anc$ on $i$ as
\begin{eqnarray}
\label{impact}
I_{\anc\rightarrow i}=\sum_{i'\in\cO_\anc\backslash
\{i\}}[fs_{ii'}^{\rm read}+(1-f)s_{ii'}^{\rm aut}]
\end{eqnarray}
with $f$ to adjust the relative weights on the two contributions.
The subsequent analysis is simplified by setting $f=0.5$ unless
otherwise specified.
We note that citations between peers \cite{wu09} do not contribute 
to the above measure, as these citations may correspond to relations
other than similarity.
For instance, if many peers rooted from parent $\anc$ cited $i$,
it implies that $\anc$ complements $i$ instead of being merely an
influential parent of $i$.
The same is true if $i$ cites many peers rooted from $\anc$,
which suggests $\anc$ being a complement of its peers instead of a 
mere influential parent.

By keeping only the reference $\alpha$ with highest $I_{\anc\rightarrow i}$ for
all $i$, we obtain a citation backbone denoted as the \simi
backbone. Cases of equal scores are extremely rare and do not
affect results (in such situations we arbitrarily choose the latest reference with highest $I_{\anc\rightarrow i}$).
In addition, we examine also the \rand
and the \lon backbone, which selects respectively
a random parent and a reference which gives rise to the longest path
to the root (most likely the latest published parent).
Other than serving as a benchmark, the \rand backbone can be
informative
as the random parent is one of the original references.
The \lon backbone instead represents a natural choice if progress is
always based on recent developments, as one may follow
the step-by-step evolution of science represented by the maximum
number of steps needed to reach the root.

\section{Statistical Properties of the Backbone}
We will examine the citation network among the journals of 
American Physical Society, from year 1893 to 2009. The dataset is
composed by $4.67\times 10^6$ citations between $4.49\times 10^5$
publications. In rare cases there are references to contemporary or
even posterior published papers. These citations are removed and the
network is strictly acyclic.

We note that all papers without reference are potentially the roots
of the backbone. As this number is in general greater than one and we
are limited to the simplest case with one selected ancestor per paper,
there may appear multiple roots and hence isolated trees in the
backbone. In the subsequent discussion, we will refer the output of
the \simi, \rand and \lon
algorithms as \emph{backbone}, and its isolated components as
\emph{trees}. Since the seemingly isolated roots may be connected by
citations other than the APS network, the number of isolated trees
would be lower if a more comprehensive citation data was used.
Nevertheless, such isolated trees may represent a crude classification
of papers. Table \ref{tab_stat} summarizes general statistical
properties of the group of trees as obtained by \simi, \rand and \lon
approaches.

\begin{table}[h!]
\footnotesize
\vspace{0.5cm}
\setlength{\tabcolsep}{7pt}
\begin{tabular}{lccc}
\hline
 & \simi &  \rand & \lon \\
\hline
Number of isolated trees & 3953 & 6594 & 2630 \\
Size: & & & \\
\qquad largest tree & $26115$ & $30358$
& $428147$\\
\qquad 2nd largest tree & $25386$ & $15697$ & $592$\\
\qquad 3rd largest tress & $21794$ & $11362$ & $471$\\
$\avg{\Delta t}$ parent-offspring & 9.5 y& 7.4 y& 2.1 y\\
\hline
\end{tabular}
\caption{Statistical properties of the isolated trees in the
\simi, \rand and \lon backbones. Values for \rand are averaged over
100 realizations. We also show the average interval (in
years) between the date of publication of a paper and its parent.}
\label{tab_stat}
\end{table}

\section{The Structure of the Backbone}

In this section, we will discuss and derive measures to validate the
citation backbones as representative of descendant charts. Three
aspects will be studied. Firstly, we examine the linkage between
different generations of papers. Secondly, we quantify the paper classification
as given by the clustering and branching structure in the backbones. 
Finally, we examine the possible self-similarity in
citation backbones.

\subsection{Hierarchy}
\label{sec_hierarchy}

\begin{figure}
\center
\includegraphics[scale=0.25]{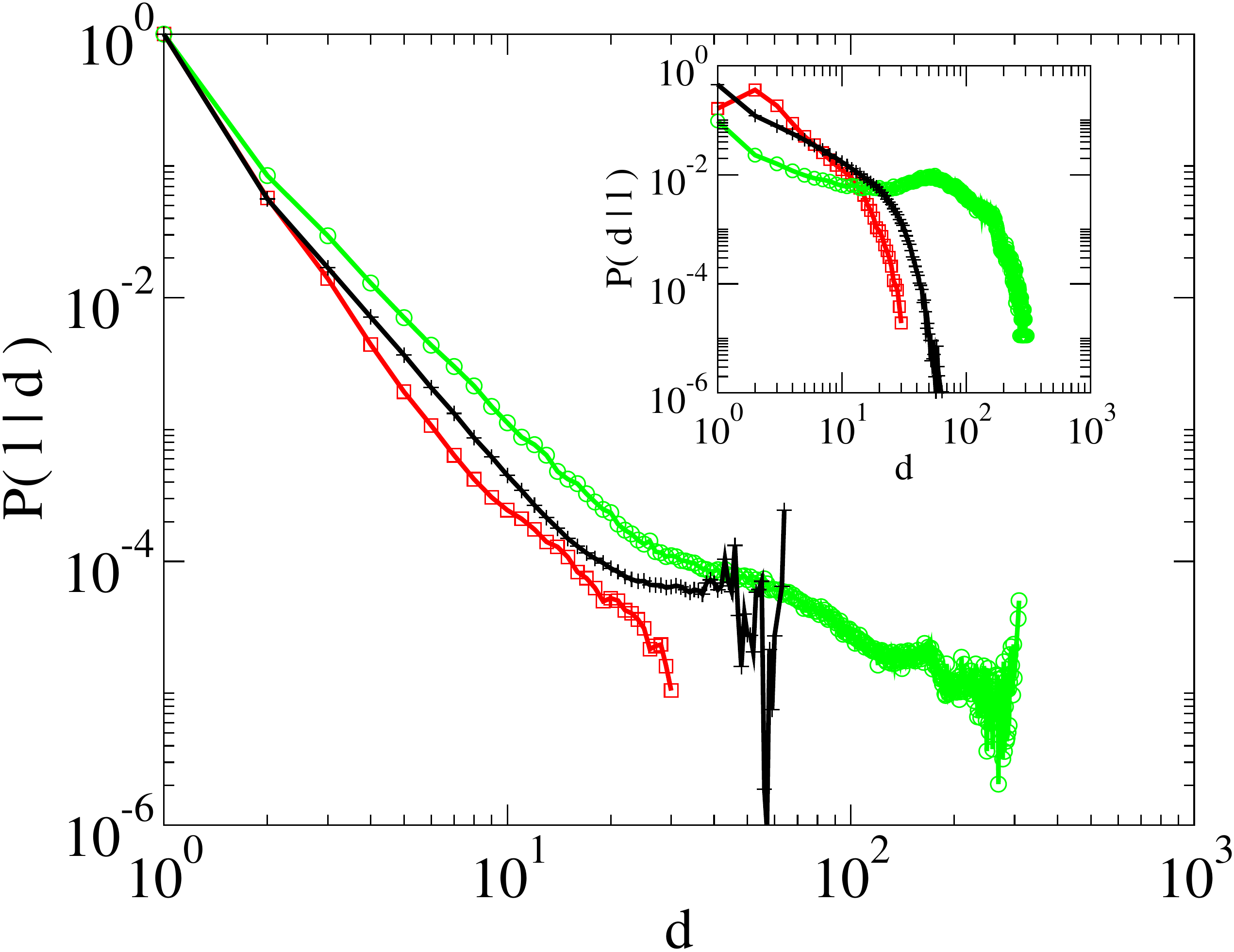}
\caption{(Color online) Conditional probability $P(l|d)$ of observing a
citation between two papers at distance $d$ on the 
\simi (red squares), the \rand (black plus) and the \lon (green circles) backbone. 
Results for \rand are averaged over
100 realizations of the backbone (the variance is negligible).
Inset: conditional probability $P(d|l)$ that two papers are
at distance $d$ given there is a citation.}
\label{PD}
\end{figure}

We first examine the probability of observing an original citation
between two papers as
a function of their distance in the backbone.
If the backbone is meaningful we expect this quantity to decrease fast
as the distance increases.  
To compute the distance between $i$ and $j$ we find the
first common ancestor $\anctwo$ in the backbone and count the
number of steps $d_{i\anctwo}$ and $d_{j\anctwo}$ 
required to go from $i$ to $\anctwo$ and from $j$ to $\anctwo$. The
distance 
$d_{ij}$ is then set as $d_{ij}=d_{i\anctwo}+d_{j\anctwo}$. We
consider $d_{ij}=\infty$ for paper $i$ and $j$ in isolated trees.
In Fig.~\ref{PD} we plot $P(l|d)$ as a function of $d$ for 
all \simi, \rand and \lon backbones, where $l$ denotes the presence of
a
link, i.e. a citation.
As we can see, $P(l|1)=1$ by definition and all $P(l|d)$ display a
power law decay for small $d$.
The \simi backbone shows a faster decay than other algorithms,
suggesting that citations are more localized in the
neighborhood of a paper in the \simi backbone.
A similar quantity $P(d|l)$ (see the inset of Fig.~\ref{PD}) also
indicates that the \simi backbone is the best representative of the
APS network since citations are concentrated at $d=2$ and decay
faster as the distance increases. 

\begin{figure}[h!]
\center
\includegraphics[scale=0.65]{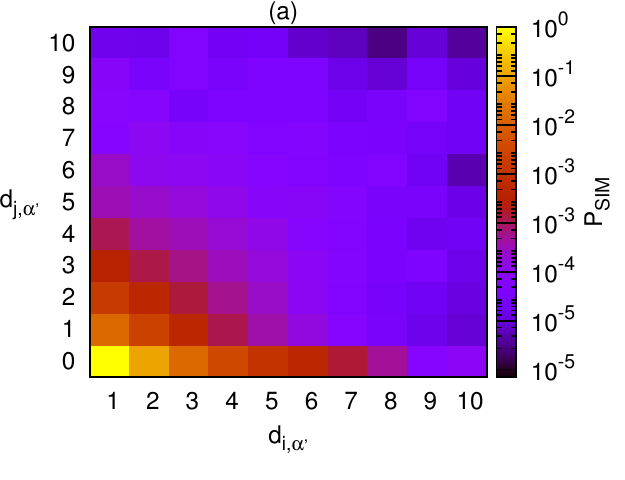}
\includegraphics[scale=0.65]{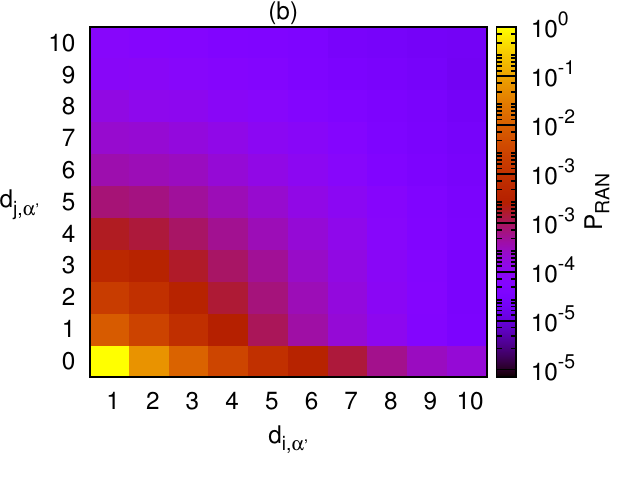}
\includegraphics[scale=0.65]{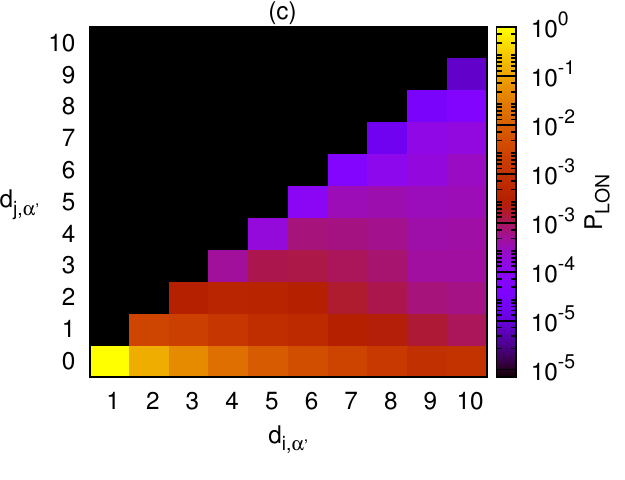}
\includegraphics[scale=0.65]{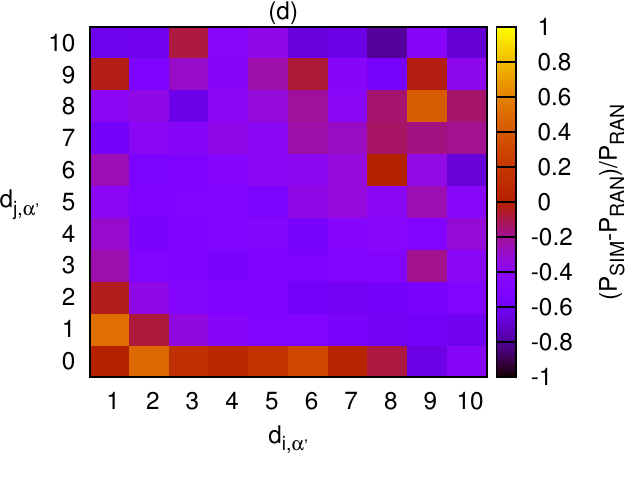}
\caption{
(Color online) Heat maps which show $P(l|d_{i\anctwo}, d_{j\anctwo})$ as a function of $d_{i\anctwo}$ and $d_{j\anctwo}$
for 
(a) the \simi backbone,
(b) the \rand backbone
and (c) the \lon backbone,
with citation $i\to j$.
Since papers only cite references published before them,
the observed dark triangle in \lon suggests a rather homogeneous temporal interval between 
papers and their best \lon ancestor, such that citation 
with $d_{j\anctwo}>d_{i\anctwo}$ are highly improbable.
Results for 
\rand are averaged over 100 realizations of the backbone (the variance
is negligible).
(d) Scaled difference of 
$P(l|d_{i\anctwo}, d_{j\anctwo})$ between \simi and \rand. 
}
\label{heat}
\end{figure}

In addition to $P(l|d)$, we further consider $P(l|d_{i\anctwo},
d_{j\anctwo})$ where $\anctwo$ is again the first common ancestor of
$i$ and $j$ in the backbone.
This allows us to see whether citations are localized on the specific
branch of each paper or spread over different ramifications on the
tree.
For any pair $(i,j)$ we take $i$ as the later published
paper such that the only potential citation is $i\to j$.
We show  in \fig{heat}~(a)-(c) the results of $P(l|d_{i\anctwo}, d_{j\anctwo})$ for the three backbones,
as a function of $d_{i\anctwo}$ and $d_{j\anctwo}$.
One notes that increasing $d_{i\anctwo}$ on the line of
$d_{j\anctwo}=0$ corresponds to the vertical trace towards the root,
while points with $d_{j\anctwo}\neq 0$ correspond to the various
`ramifications' in the backbone.
Both \simi and \rand gives a meaningful structure where citations are
localized on the descendant chart of the immediate and next immediate
ancestor, i.e. the triangle in the bottom left-hand corner. Citations
between different ramifications are rare.
The \lon backbone instead displays a less coherent structure
where citations crossing different lines of research are common.
To examine the difference between \simi and \rand we also show the
scaled difference of their $P(l|d_{i\anctwo}, d_{j\anctwo})$ as given in the
vertical axis of \fig{heat}~(d). This
comparison clearly indicates that \simi gives raise to the most
meaningful hierarchy as citations are mainly found on the descendant
chart of the more relevant ancestors instead of crossing different
charts. 

\subsection{Clustering}
\label{sec_clustering}

In addition to the crude classification as given by the isolated
trees, the branches in a single tree are also informative to identify
research fields and sub-fields.
From the clustering point of view the method we have introduced is
computationally efficient (with complexity to be
$O(N)$ as long as connectivity is not extensive) compared
to modularity maximization based algorithms \cite{newman04,newman06}
or hierarchical clustering algorithms \cite{jain99}
(with complexity at least $O(N^2)$).
Moreover, the clustering naturally explores the temporal
dimension by preserving the ancestor-descendant relations. 

In order to map the backbone into clusters we consider two
simple approaches which involve only a single parameter.
The first approach makes use of the publication year of papers and
naturally follows the order of publication.
We first make a cut at the year $Y_c$ such that papers printed before
$Y_c$ are removed. We then consider each
unconnected component as a different branch, i.e. a different cluster, in the original backbone,
and as a classification for papers.

The second approach is dependent on the cluster size which we consider
to be a typical research branch.
Starting from the leaves of the backbone (i.e. papers with no
offspring) we trace towards the root until a branching point is
reached. The branching point is defined as a node of the network
from which at least (i) two ramifications start and (ii)
two ramifications are extended more than $S$ steps. 
When a branching point satisfies these requirements,
all ramifications originating from it are considered as different
branches, resulting in a classification of papers.
Here we quantify the validity of clustering as a function of parameter
$Y_c$ and $S$.

In order to evaluate the quality of a given clustering we use two
different measures.
The first one---which we call \emph{exclusivity}---is a 
modified modularity measure specific for directed
acyclic graphs.
The rationale behind this measure is to compute the fraction of links
of the original network falling inside the same cluster and compare it
with the expected value for a random directed acyclic graph.
We denote the set of papers assigned to branch $x$ as $\cX$ and
define the \emph{exclusivity} as
\begin{equation}
E = \left\langle\left\langle \frac{p^i_x}{p_i}-\frac{n^i_{x}}{n_i}
\right\rangle_{i\in \cX}\right\rangle_x
\end{equation}
where $p_i$ is again the number 
of references of $i$; $p^i_x$ is the number of $i$'s references in
branch $x$; $n_i$ is the number of papers published before $i$;
$n^i_x$ is the number of papers published before $i$ in branch $x$.
The term $n^i_x/n_i$ thus corresponds to the expected fraction of
links from $i$ to an element in $\cX$ in the random case. To reduce
the noise from small clusters, we have excluded branches with less
than 10 papers.

The second measure we use is the effective number
of PACS---$N_P$---which counts the average number of heterogeneous
PACS in individual branches. Good paper classifications
result in small values of $N_P$.
We first denote $r^x_p$ to be the fraction of paper in branch $x$
which
is labeled by the PACS number $p$, 
and note that $\sum_p {r^x_p}\ge 1$ as papers are always 
labeled by more than one PACS number.
$N_P$ is then defined as
\begin{equation}
\label{eq_E}
N_P =  
\left\langle\frac{1}{\sum_p
(f^x_p)^2}\right\rangle_{x},
\end{equation}
where $f^x_p=r^x_p/\sum_{p'}r^x_{p'}$. Therefore, $N_P=1$ when there
is only one PACS in the branch
which corresponds to the optimal classification of papers. On the
other hand, $N_P$ attains its maximum when 
all PACS numbers in $\cX$ have equal share (i.e. equal $f^x_p$) and a
large $N_P$ thus corresponds to high heterogeneity inside single
clusters. We remark that in evaluating $N_P$, only the first four
digits are used to distinguish PACS number.

\begin{figure}
\vspace{10pt}
\centering
\includegraphics[scale=0.165]{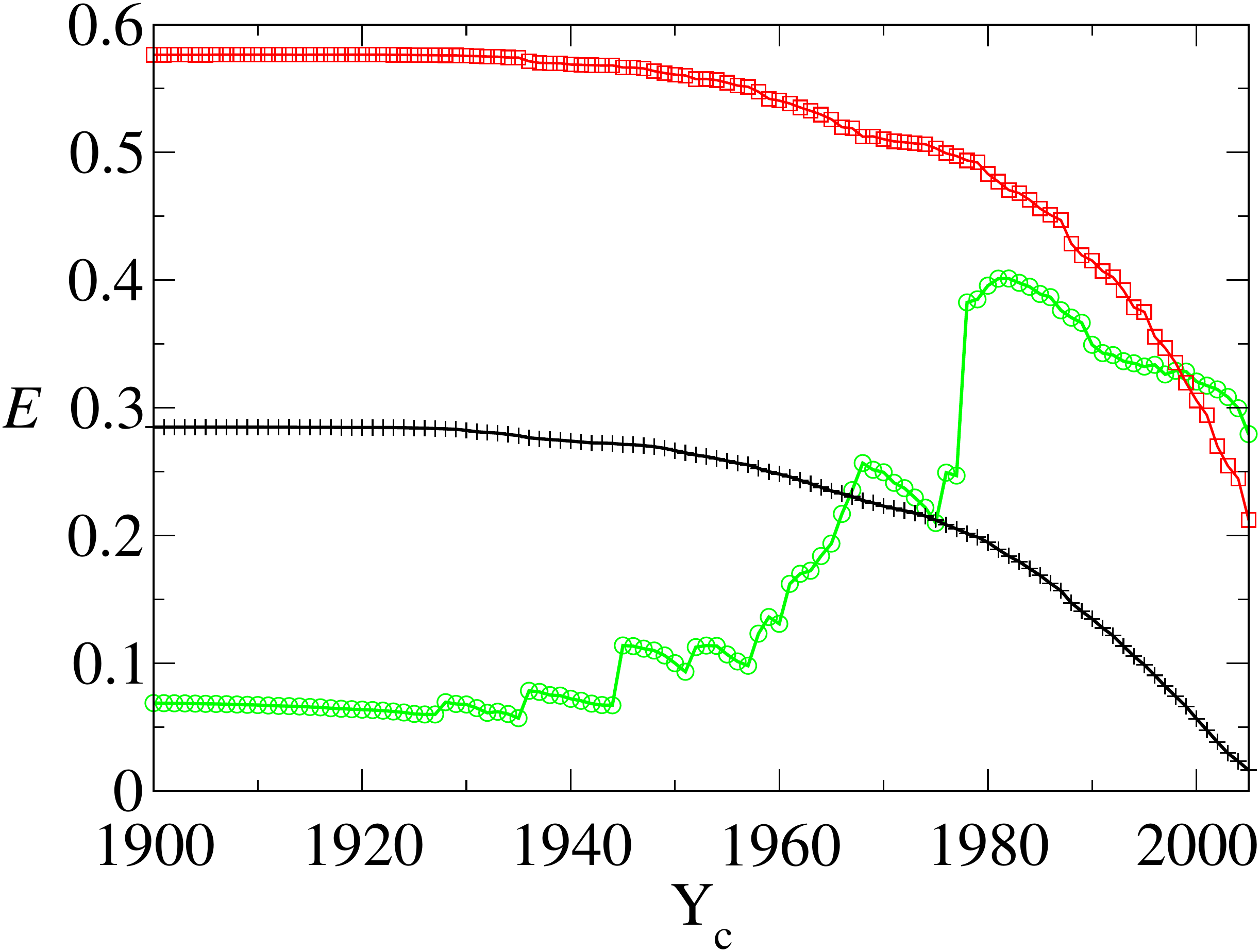}
\includegraphics[scale=0.165]{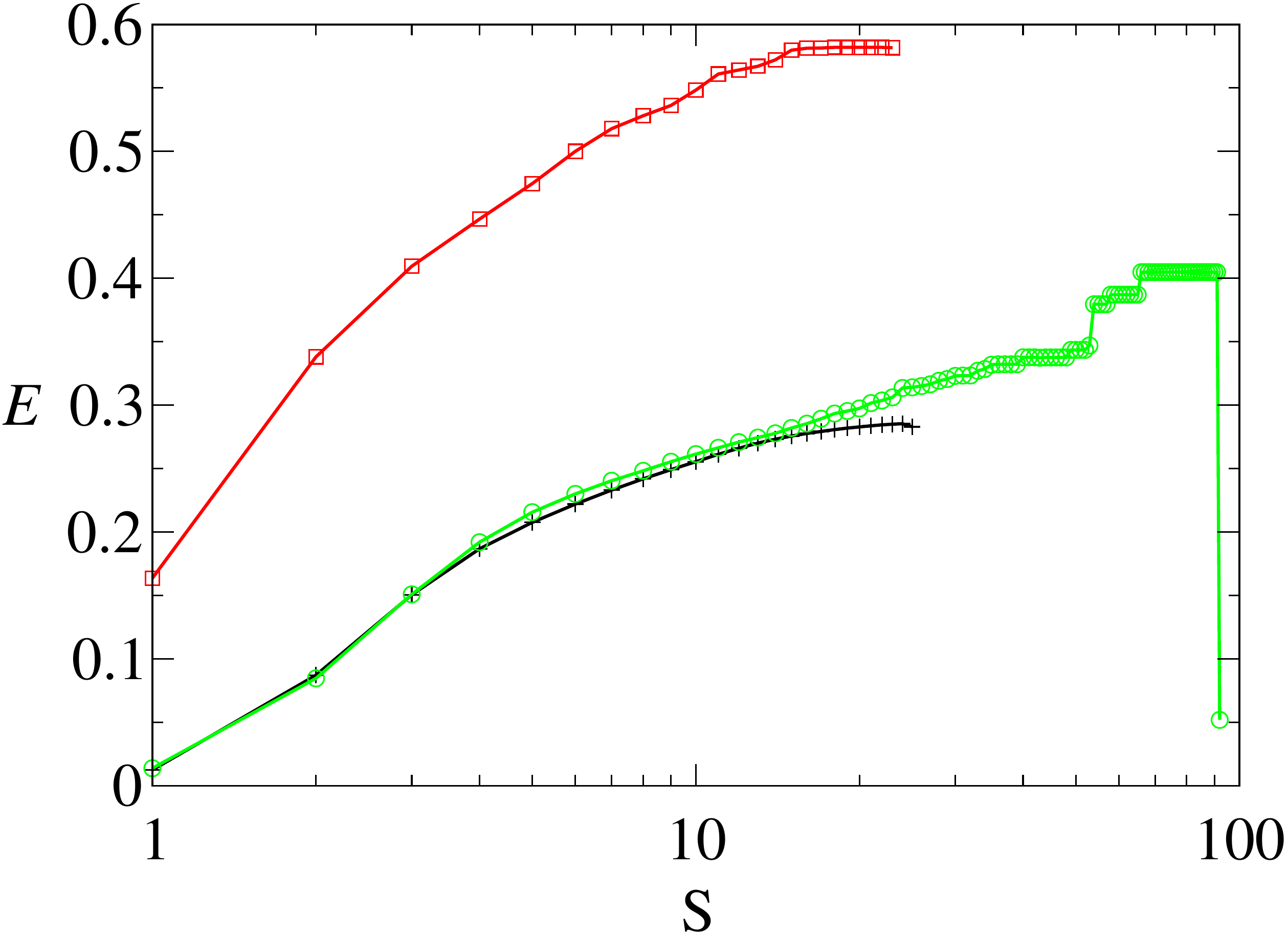}
\includegraphics[scale=0.165]{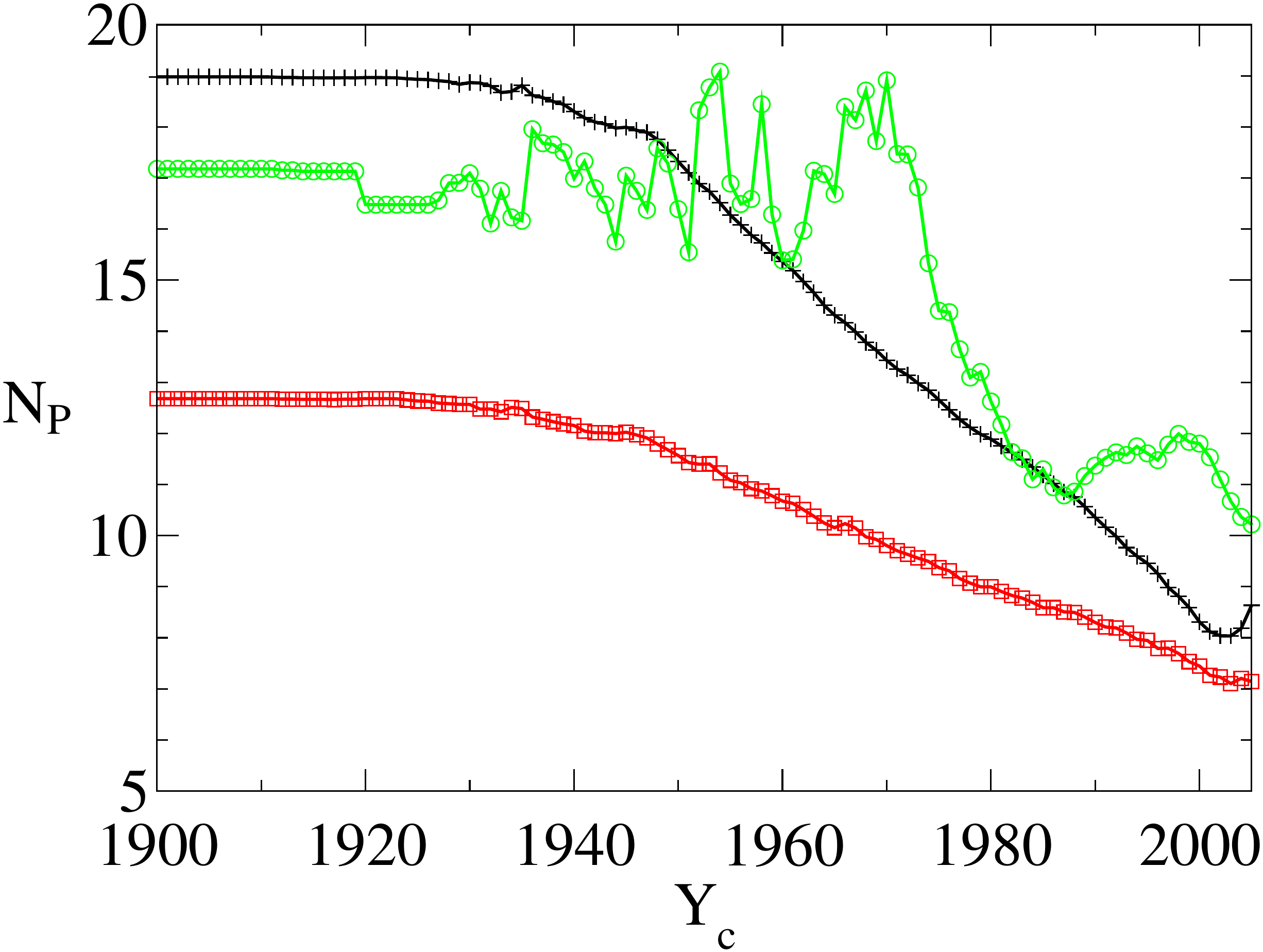}
\includegraphics[scale=0.165]{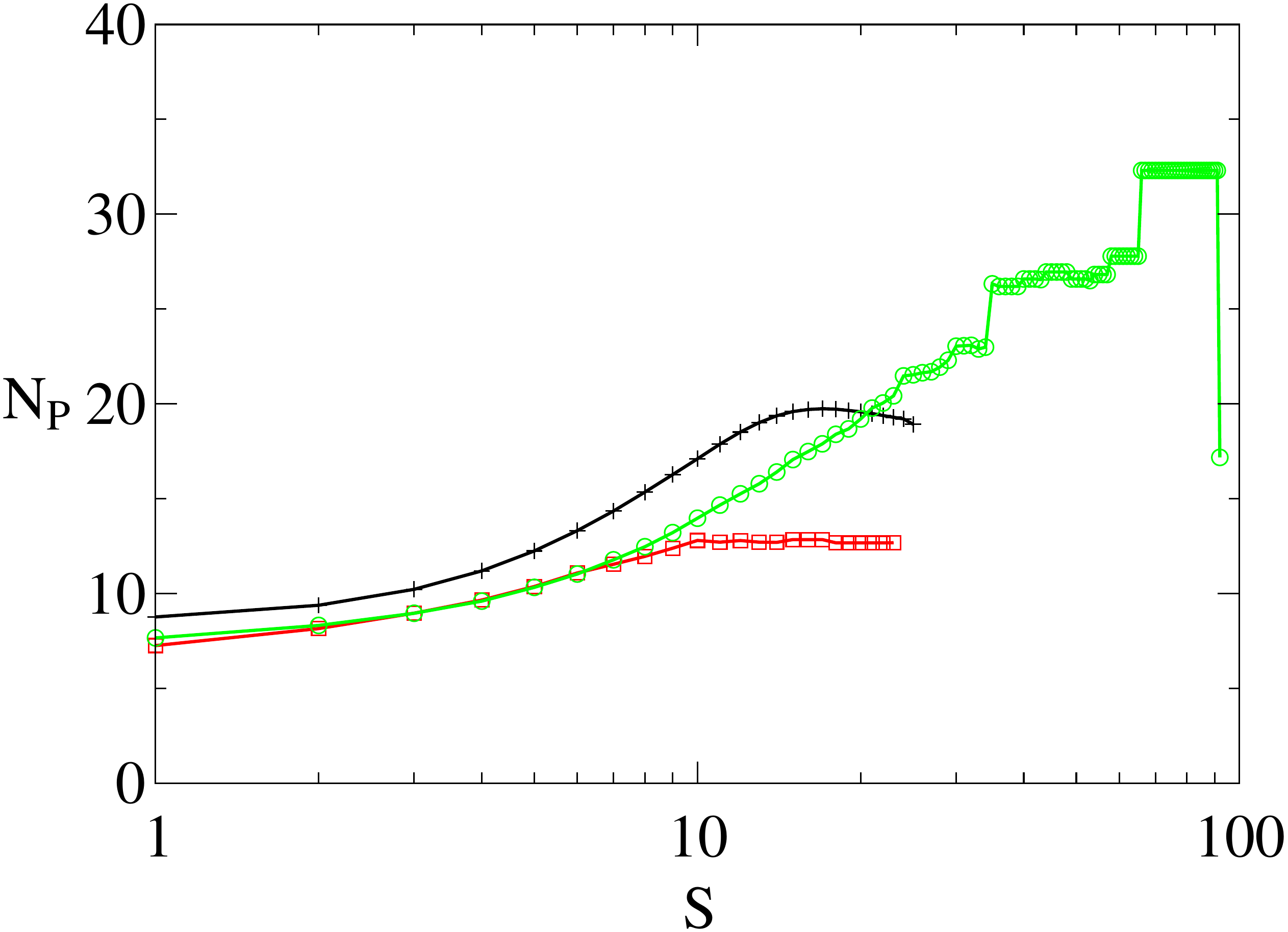}
\caption{(Color online) The exclusivity $E$ and the effective number
of PACS
$N_P$ as a function of cut-year $Y_c$ and branch depth $S$ for the 
\simi (red squares), the \rand (black plus) and the \lon (green circles) backbone. 
Both quantities show that $\emph{SIM}$ gives a more meaningful
division into branches. Results for \rand are averaged over 100
realizations of the backbone (the variance is negligible).}
\label{fig:clustering}
\end{figure}

In Fig.~\ref{fig:clustering} we plot the 
$E$ and $N_p$ as a function of the two parameters $Y_c$ and $S$. 
Both measures are biased by the cluster size but in an opposite way.
While $N_p$ indicates better clustering (and thus a lower value) when
isolated clusters are of smaller size, $E$ indicates better clustering
(and thus a higher value) when clusters are of larger size. Even with
the compensation by $n^x_i/n_i$ in \req{eq_E}, we still observe a
small bias of $E$ on cluster size. These biases may influence our
comparison of the identified clusters from the \simi, \rand and \lon
backbones, as they have different sizes. Nevertheless,
the combination of the two independent measures clearly indicate that
\simi is the best choice to obtain a meaningful clustering besides the
bias introduced by cluster sizes. Moreover the exclusivity of the
\simi backbone is higher for any value of the parameter $S$ which 
further supports the validity of the comparison despite the presence
of the bias.

\subsection{Self-similarity}

Other than the hierarchical and clustering properties,
the backbones may possess self-similarity.
Intuitively, self-similarity may be induced when branches of research 
successively generate branches of significant advances.
The existence of fractality in the backbone would provide support 
for its relevance with the evolution of science.

To show the self-similarity in networks,
one can measure their fractal dimension by the box-covering method 
\cite{song05, goh06, kim07}.
In this approach, the fractal dimension $d$ is defined as the
power-law exponent in
\begin{eqnarray}
N(l_B) \sim l_B^d,
\end{eqnarray}
where $N(l_B)$ is the minimum number of boxes, each of radius $l_B$,
required to cover the whole network.
To obtain the exact $N(l_B)$ is in general difficult,
we thus employ the random sequential box-covering algorithm
\cite{kim07}
which gives an approximate $N(l_B)$ with the same scaling.
Specifically, we start with all nodes being ``uncovered" and 
repeat the following procedures until all nodes become ``covered'':
(1) pick randomly a seed node,
(2) find all ``uncovered'' nodes within a distance of $l_B$ from the 
seed, and
(3) increase $N(l_B)$ by one if there exists at least one
``uncovered'' node and mark all of them as ``covered''.
Note that a ``covered'' node can also be a seed in the subsequent
searches. For the same tree,
we show the minimum of $N(l_B)$ among 20 random sequences
as our final value for each value of $l_B$.

\begin{figure}
\includegraphics[scale=0.25]{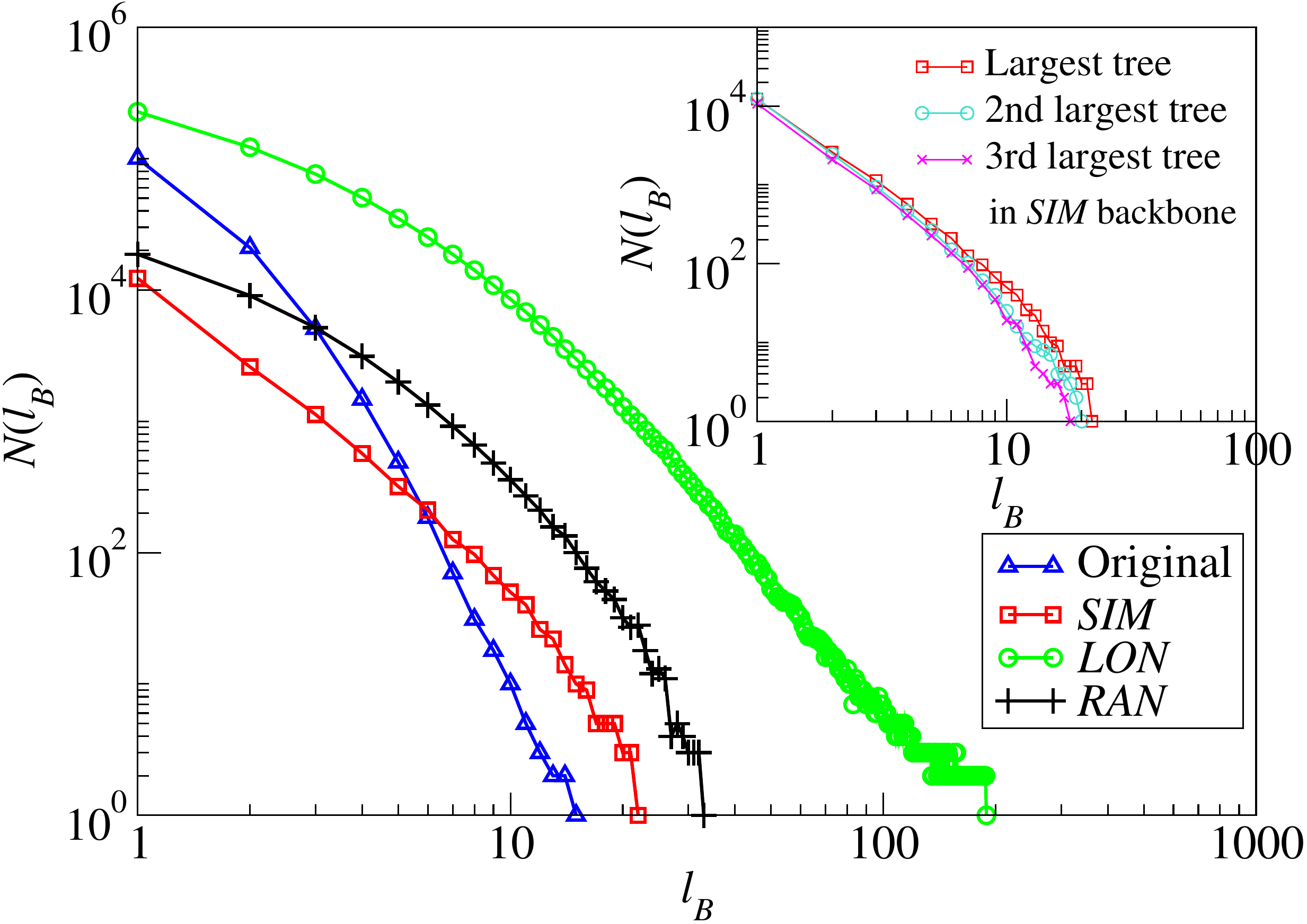}
\caption{(Color online)
$N(l_B)$ as a function of $l_B$ for the largest tree in 
the original network and its \simi, \lon and \rand backbone,
taken as the minimum over 20 random sequences of seed nodes for the
box-covering
algorithms.
The \emph{SIM} backbone are obtained at $f=0.5$.
Inset: $N(l_B)$ as a function of $l_B$ for the three largest trees in
the \simi backbone.
}
\label{fig_box}
\end{figure}

We show in \fig{fig_box} the results of $N(l_B)$ as a function of
$l_B$ for the largest tree 
in the \simi, \rand and \lon backbone.
The results are compared to $N(l_B)$ of the original citation
network.
As we can see, $N(l_B)$ from the
\lon backbone has the highest resemblance to power-laws, while that
of the \rand backbone shows the fastest decay in $N(l_B)$.
The \lon tree has a long tail of $N(l_B)$, 
as it is longest and largest in size (see \tab{tab_stat}).
Only the largest tree of a particular realization of the \rand
backbone is shown, as similar results are observed in other realizations.
Though a long tail is not observed in the \simi tree,
it shows a power-law-like behavior up to an intermediate value of
$l_B$.
Similar behaviors are also observed in the other isolated trees of the
\simi backbone,
as shown by the inset of \fig{fig_box}.

We interpret the results as follows.
The observed resemblance to power-laws from the \simi and \lon backbone may suggest the
presence of self-similarity in their descendant chart.
While the \lon backbone does not possess a meaningful hierarchy or
clustering compared to the \simi backbone,
its step-by-step structure indeed shows the highest fractality.
We note that a rather short power-law is also observed in the original
network, though characterized by a different exponent from the \simi and \lon
backbones.
On the other hand, such fractality is not observed in the \rand
backbone.

\section{Potential Applications}

In this section, we briefly describe the implications and
potential applications of the citation backbone as a descendant chart of research papers.

As the backbone is a sketch of the skeleton of scientific development,
it can be applied to identify seminal papers.
Preliminary results show that a simple measure based on the
the number of \emph{relevant offspring}, i.e. followers in the
backbone, is sufficient to give a meaningful ranking
that is not trivially correlated with the original number of
incoming citations (between the two ranking the Kendall's correlation
coefficient is $0.19$ and there is an overlap of only 7 papers in the
top 20 ranks). This serves as a simple yet meaningful definition of
impact of a publication.
More refined definitions which takes into account
the reputation of each relevant offspring and/or the structural role
of a given paper in the backbone can give even better selection of
fundamental papers.
Moreover, our formulation of tunable weight on authors' and readers'
interpretation in \req{impact} can be easily incorporated 
in common ranking algorithms such as Page Rank where an even
repartition of citation importance is instead assumed.

The second application corresponds to the classification of papers. As
we have mentioned before, such clustering divides papers into research
fields or sub-fields and offers a basis for a synthetic picture of
the state-of-the-art. 
There are several advantages over conventional classifications, which
include (1) lower computational complexity, (2) additional information
of sub-clustering as given by the internal tree structure, (3)
predictions of future development by considering the rate of growth of
sub-branches.
Especially this last feature is useful to filter the most active
directions in the large amount of literature at our disposal.

\section{Conclusions}

We have shown that a simple backbone constructed by the
most relevant citations can well characterize the original citation
network.
Conversely, non-trivial informations stored in the citation network
can be simply extracted from its backbone.
While conventional spanning trees are based on contemporary
information, we demonstrated the significance of temporal dimension in
citation backbones.

Specifically, we have introduced both a simple approach to
identify the most relevant reference for each publication and
effective measures to quantify the validity of the resulting
backbone. 
Our results show that the essential features of hierarchy and paper clustering in the original network
are well captured by our citation backbone,
while this is not the case for other simple approaches.
On the other hand,
we showed that resemblance to self-similarity is observed in citation backbones.

In terms of applications, the backbone can be considered as a
descendant chart of research papers, which constitutes a useful basis
for identifying seminal papers and paper clusters, and
in general a synthetic picture of different research fields.
In particular, paper classification by mean of the backbone is
computationally efficient when compared to the conventional clustering
approaches, and provides additional information on the cluster
structure besides a mere cluster label.

While we only investigated the citation network of the American
Physical Society, the same approach can be readily applied to other citation
networks. It would be also interesting to examine the
potentials of the present approach on other directed acyclic graphs.

\section*{Acknowledgements}

We thank Mat\'u\v s Medo for fruitful discussions and comments,
Giulio Cimini and An Zeng for meaningful suggestions.
This work is supported by QLectives projects (EU FET-Open
Grants 231200), the Swiss National Science Foundation
(Grant No. 200020-132253), the Sichuan Provincial Science and
Technology Department (Grant No. 2010HH0002) and the National Natural
Science Foundation of China (Grant Nos. 90924011,60973069).
CHY is partially supported by EU FET FP7 project STAMINA (FP7-265496).
We are grateful to the APS for providing us the dataset.

\end{document}